\documentstyle[psfig]{caosp}
\begin{document}
\def\Hm{\langle H\rangle}
\def\Hs{(\Hsq+\Hzsq)}
\def\Ht{\Hs^{1/2}}
\def\Hsq{\langle H^2\rangle}
\def\Hzsq{\langle H_z^2\rangle}
\hauthor{S. Hubrig {\it et al.}}
\title{Magnetic Bp and Ap stars in the H-R diagram}
\author{S. Hubrig \inst{1} \and P. North
\inst{2}   \and G. Mathys \inst{3}}
\institute{University of Potsdam, Am Neuen Palais 10, D-14469 Potsdam, Germany
\and Institut d'Astronomie de l'Universit\'e de Lausanne,
CH-1290 Chavannes-des-Bois, Switzerland
\and European Southern Observatory, Casilla 19001, Santiago 19, Chile}

\date{\today}
\maketitle
\begin{abstract}
The positions in the H-R diagram of strongly magnetic Ap and Bp
stars are compared with those of normal main sequence stars of
types B7 to F2, with a view to investigating possible differences in
evolutionary status between magnetic and non-magnetic stars.
The normal B7--F2 stars fill the whole width
of the main sequence band with some concentration towards the ZAMS,
whereas the magnetic stars are only rarely found close to either the
zero-age or terminal-age sequences.
\end{abstract}
\section{Introduction}
\label{intr}
The evolutionary status of Bp and Ap stars was hardly settled in the past.
Estimates on their evolution were based on the membership of magnetic stars in
open clusters or associations, on their membership in binary systems, or
on indirect arguments inferred from the assumption of a rigid rotator
geometry.
More recently, it was advocated that Ap and Bp stars are distributed
uniformly across the width of the main sequence (North 1993; Wade 1997),
or alternatively that
the magnetic stars are near the end of their main sequence life
(Hubrig \&\ Schwan 1991; Hubrig \&\ Mathys 1994; Wade et al. 1996).
One open question was  whether part of the apparent
inconsistency between these results might be related to the fact that
not all Ap and Bp stars are necessarily strongly magnetic.


Now, with the release of the Hipparcos data, it has become possible to
determine the evolutionary state of magnetic stars with more reliability.
In order to understand the physical processes taking
place in B and A stars, we investigated possible differences of
evolutionary state between magnetic and non-magnetic stars.

\section{Basic data}
\subsection{Sample of magnetic stars}
For the present study,
Hipparcos parallaxes and photometric data
have been compiled exclusively for two groups of stars with strong, 
well-established magnetic fields on their surface. The first group consists of
stars known to have spectral lines resolved into their
magnetically split components when observed in unpolarized light
(Mathys et al. 1997).
The second group includes  Bp and Ap stars for which the mean 
quadratic magnetic field has been determined through application of the moment
technique (Mathys 1988; Mathys 1995; Mathys \&\ Hubrig 1997).
While the first group is strongly biased
against rapidly rotating objects, the second one is not affected by this bias.

Accurate Hipparcos parallaxes ($\sigma(\pi)/\pi<0.2$) are available
only for 23 stars with magnetically resolved lines and for 12 stars with
known mean quadratic magnetic fields. Thus the whole sample contains
35 magnetic stars.

Effective temperatures of the magnetic stars have been derived from
$uvby\beta$ data or from photometric data in the Geneva system. In almost all
cases these temperatures are in good agreement with the spectral type
classification taken from the literature. For 5 hot stars with mean
quadratic field determinations, the discrepancy is rather large, up to
a few thousand Kelvin. For these stars, effective temperatures were inferred
from detailed spectroscopic studies available in
the literature or from the spectral types.

\subsection{Sample of normal B and A stars}
In order
to compare the positions of strongly magnetic Ap and Bp stars in the H-R
diagram with those of normal main sequence stars, we selected
all normal single main-sequence stars of spectral types B7 to
F2 from the BS catalogue, with accurate Hipparcos parallaxes. This sample
consists of a total of 416 stars at distances below 100 pc ($\pi>10$ mas).

For all stars in both
samples, we applied corrections for IS absorption and duplicity (for magnetic
stars), and also the LK correction (Lutz \&\ Kelker 1973).

\section{Analysis and results}

We find that the normal B7--F2 stars fill the whole width
of the main sequence band with some concentration towards the ZAMS.
Magnetic stars are only rarely
found close to either the zero-age or terminal-age sequences.

No clear evidence of an evolution
of magnetic field strengths across the main sequence is found. But we note
that the stars with the strongest magnetic fields in most cases
appear in the middle of the main-sequence band. This implies that
they should already have
spent a considerable fraction of their life on the main sequence.
A possible interpretation of the distribution of magnetic stars in
the H-R diagram might be that magnetic stars represent a transitory phase in 
the evolution of normal stars across the main sequence.
Since the region occupied by non-magnetic stars in the H-R diagram
fully overlaps that where the magnetic stars are found, it would then appear
that not all B and A stars pass through a ``peculiar'' phase.

A Kolmogorov-Smirnov test
applied between the distributions of the normal stars and of each group of
magnetic stars has shown that the distribution of the values of $\log g$ for
the sample of stars with magnetically resolved lines differs from the
distribution for non-magnetic stars at a significance level of
99.2\%. For the whole sample of magnetic stars,
the difference with the non-magnetic stars is significant at the
92.6\%\ level.

\section{Discussion}
In order to test the theories of magnetic field origin, it would be important
to probe the evolution of the magnetic field strength across the main
sequence.
Our sample of magnetic stars is too small to provide a really
stringent test. The sample could be only marginally enlarged
by incorporating
stars with strong longitudinal fields and accurate Hipparcos parallaxes.
Another relevant issue that should be considered is
the evolutionary state of chemically peculiar B and A stars
without detectable or with very weak magnetic fields. 

The study of the magnetic field geometry in stars
of various ages and rotation rates will provide important clues
to test theoretical predictions. Several
mechanisms have been proposed
by which the angle between magnetic and rotation axes
might change during the main-sequence life time. Therefore
one goal for
observers should be to provide theorists with constraints
on the distribution of magnetic field geometries.

\end{document}